\documentclass[10pt,aps,pre,noshowpacs,twocolumn,superscriptaddress,nobibnotes,longbibliography]{revtex4-1}
\usepackage{amssymb}
\usepackage{graphicx}
\usepackage{amsmath,amsfonts}
\usepackage[colorlinks=true,allcolors=blue]{hyperref}
\usepackage[utf8]{inputenc}
\usepackage{xcolor}
\usepackage[normalem]{ulem}
\usepackage[T1]{fontenc} 
\usepackage{multirow}
\usepackage{booktabs}
\usepackage{threeparttable}
\usepackage{tabularx}
\usepackage{float}
\usepackage{wasysym}
\usepackage{harmony}

\definecolor{dartmouthgreen}{rgb}{0.05, 0.5, 0.06}

\definecolor{randomcolor}{rgb}{0.55, 0.05, 0.06}

\usepackage{pdfpages}
\makeatletter
\AtBeginDocument{\let\LS@rot\@undefined}
\makeatother

\begin{document}

\title{Trade-offs between structural richness and perceptual robustness in\\ music network representations}
\date{\today}

\author{Lluc Bono Rosselló}
\email[Corresponding author: ]{Lluc.Bono.Rossello@ulb.be}
\affiliation{Institute for Interdisciplinary Studies on Artificial Intelligence (IRIDIA), Université Libre de Bruxelles, Brussels, Belgium}
\author{Robert Jankowski}
%\email[Corresponding author: ]{research@robertjankowski.net}
\affiliation{Faculty of Electrical Engineering, Mathematics and Computer Science, Delft University of Technology, 2628 CD, Delft, Netherlands}
\affiliation{Departament de F\'isica de la Mat\`eria Condensada, Universitat de Barcelona, Mart\'i i Franqu\`es 1, E-08028 Barcelona, Spain}
\affiliation{Universitat de Barcelona Institute of Complex Systems (UBICS), Universitat de Barcelona, Barcelona, Spain}
\author{Hugues Bersini}
\affiliation{Institute for Interdisciplinary Studies on Artificial Intelligence (IRIDIA), Université Libre de Bruxelles, Brussels, Belgium}
\author{Mari\'an Bogu\~{n}\'a}
\affiliation{Departament de F\'isica de la Mat\`eria Condensada, Universitat de Barcelona, Mart\'i i Franqu\`es 1, E-08028 Barcelona, Spain}
\affiliation{Universitat de Barcelona Institute of Complex Systems (UBICS), Universitat de Barcelona, Barcelona, Spain}
\author{M. {\'A}ngeles Serrano}
% \email{marian.serrano@ub.edu}
\affiliation{Departament de F\'isica de la Mat\`eria Condensada, Universitat de Barcelona, Mart\'i i Franqu\`es 1, E-08028 Barcelona, Spain}
\affiliation{Universitat de Barcelona Institute of Complex Systems (UBICS), Universitat de Barcelona, Barcelona, Spain}
\affiliation{ICREA, Passeig Llu\'is Companys 23, E-08010 Barcelona, Spain}

\begin{abstract}
Music is a structured and perceptually rich sequence of sounds in time, whose perception is shaped by the interplay of expectation and uncertainty about what comes next.
Yet the uncertainty we infer from music depends on how the musical piece is encoded as an event sequence.
In this work, we use network representations, in which event types are nodes and observed transitions are directed edges, to compare how different feature encodings shape the transition structure we recover and how robust that structure is under modeled perceptual constraints.
We systematically analyse eight encodings of piano music, from single-feature vocabularies to richer multi-feature combinations. These representational choices reorganize the state space and fundamentally reshape network topology, shifting how uncertainty is distributed across transitions.
To connect these descriptive differences to perception, we adopt a perceptual-constraint model that captures imperfect access to transition statistics. Overall, compressed single-feature representations yield dense transition structures with higher entropy rates, corresponding to higher average uncertainty per step, yet low model error, indicating that the constrained estimate stays close to the corpus transitions. In contrast, richer multi-feature representations preserve finer distinctions but expand the state space, sharpen transition profiles, lower entropy rates, and increase model error.
Finally, across representations, uncertainty concentrates in diffusion-central nodes while model error remains low there, suggesting an informational landscape in which predictable flow coexists with localized surprise.
Overall, our results show that feature choice shapes not only the networks we reconstruct, but also which transition statistics are available and how vulnerable they are to distortion under perceptual constraints.
\end{abstract}

\maketitle
\let\oldaddcontentsline\addcontentsline% Store \addcontentsline
\renewcommand{\addcontentsline}[3]{}% Make \addcontentsline a no-op

\section{Introduction}
Music is a highly complex phenomenon composed of low-level features that combine into patterns that incorporate multiple structural aspects. Gaining insight into how these patterns emerge from basic musical events offers a useful perspective for understanding both the complexity of music and the perceptual processes involved in listening~\cite{koelsch2014brain,koelsch2019predictive}.
These perceptual processes include the formation of expectations about what comes next and the experience of varying uncertainty and surprise as those expectations are confirmed or violated~\cite{koelsch2019predictive,cheung2019uncertainty,mas2025predictive}.

Any analysis of musical structure depends on how the musical piece is encoded as a sequence of events. Musical sequences can be represented at different levels of detail by selecting subsets of features—such as pitch, interval, or duration—that preserve some distinctions while discarding others, an idea closely related to \textit{viewpoint}-based approaches in music cognition~\cite{conklin1995multiple}. 
These representational choices define the event vocabulary and its granularity, and therefore shape which regularities can be expressed or detected, including uncertainty about what comes next. Accordingly, what matters is not only the overall level of this uncertainty, but also how it is organized in a piece~\cite{koelsch2019predictive,cheung2019uncertainty}. Since both depend on the chosen event vocabulary, representational choices can systematically reshape the uncertainty patterns we infer.

For example, consider a short melody of eight notes: (A4–C5–A5–F5–A6–C4–F4–A7). If we encode each note as a distinct event by combining pitch class and octave, the sequence contains eight different event types, so each event type tends to have only one observed next-event option, making transitions appear highly predictable. If, instead, we collapse octave and keep only pitch class, the same melody reduces to the pitch-class sequence (A–C–A–F–A–C–F–A) and to just three event types (A, C, and F) repeated across the eight notes. Each event type can then be followed by different events (e.g., A can be followed by either C or F), increasing uncertainty. Thus, although both encodings describe the same musical surface, they induce different transition statistics, raising a basic methodological question: how does representational detail shape the structure we infer from music?

One way to make such differences explicit is to use network representations, in which sequence elements are modeled as nodes and observed transitions between elements as directed edges.
Previous work used pitch-based representations with octave and duration~\cite{liu_tse_small_2009,ferretti2017}, pitch-class sets~\cite{gomez_lorimer_stoop_2014,serrà_corral_boguñá_haro_arcos_2012,nardelli_tonal_2021,manaris2005zipf,serra2021heaps}, and chord-split representations~\cite{kulkarni2024information,di2025decoding}, which resulted in different topologies and statistical insights~\cite{del2008universality,zanette2006zipf,moss2019statistical,mehr2019universality,rohrmeier2008statistical}. 
Yet, without a systematic comparison across levels of detail within the same musical corpus, it remains unclear whether the reported topological patterns reflect the music itself or the encoding, motivating a cross-representation analysis that makes explicit how different event vocabularies redistribute uncertainty across the network.

Network representations allow us to describe how different encodings redistribute uncertainty. However, because expectation and surprise are perceptual phenomena, the relevant question is not only what uncertainty is present in the empirical transition structure, but also how reliably that structure can be internalized under imperfect memory and noise.
Musical expectation is shaped by exposure: listeners learn statistical regularities from musical experience, and these learned regularities influence what later sounds expected or surprising~\cite{huron2008sweet,pearce2018statistical}. 
If such transition statistics are not internalized perfectly, the structure available for prediction may be a biased approximation of the empirical sequence, and later expectations would inherit this distortion.

This raises a second question: under what conditions does a representation's transition structure remain stable when it is internalized under perceptual constraints, and therefore provide reliable statistics for later expectation formation?
Human responses to sequential input are limited by memory, noise, and representational cost~\cite{lynn2020human, lynn2020abstract,momennejad2017successor,howard2002distributed,meyniel2016human}. A key consequence is that observers may not rely only on the immediately observed one-step transitions, but may partially integrate information over multiple steps—treating events that are reachable in two or more steps as if they were more directly available than the true statistics imply. Such integration can cause listeners to treat events that occur only after several steps as if they were possible next events, thereby changing the transition probabilities they infer. Recent work shows that the recoverability of transition statistics under such perceptual constraints depends on the organization of uncertainty~\cite{lynn2020human}. Lynn et al.~\cite{lynn2020abstract} showed, using reaction-time experiments with visual sequences, that even when two sequences have the same average uncertainty, meaning that continuations are equally unpredictable, differences in the organization of transitions can cause humans to react as if the uncertainty were higher, resulting in a perceptual error,  defined as the mismatch between the true transition statistics and the observer’s inferred model.

Returning to our melodic example, this suggests that the pitch-class+octave encoding should allow for a single empirical continuation at each step, yet a learner with imperfect access to the sequence may still fail to recover this structure and treat other continuations as plausible. For instance, after observing an event such as A6, the true continuation in the sequence is C4. However, a biased estimate may also assign probability to an alternative, such as F4, consistent with the multi-step integration described above, in which events reachable after multiple steps are treated as if they were directly available next. Introducing such competing continuations increases the uncertainty of the internalized transition structure and alters the inferred transition structure from a single-outgoing edge pattern to a branching one. Conversely, in the pitch-class-only representation, the same context reduces to A, for which both continuations C and F are in fact observed in the melody (i.e., A is followed by C and also by F). In that case, an error that mixes up which continuation occurred does not introduce a qualitatively new transition, but mainly shifts probability mass between already-available options (e.g., between A$\to$C and A$\to$F), making the representation more robust to such distortion during internalization. Therefore, understanding how topology modulates this distortion of transition statistics is necessary for evaluating which representations provide transition structures that are more stable under perceptual constraints and may therefore serve as more reliable inputs for later prediction.

In short, the choice of representation fixes the transition statistics and therefore the distribution of uncertainty. Representing these statistics as a network makes their topology explicit. Because humans do not access transition statistics perfectly, perceptual constraints can make the same statistical uncertainty appear higher or lower, depending on how transitions are organized. This motivates (i) a systematic comparison of network structure across levels of representation detail and (ii) an analysis of how the resulting network topology relates distortion under perceptual constraints.
 
In this work, we address this issue by analysing eight network representations of music, drawing on and extending existing approaches in the literature. These vary in their level of detail, from single-feature encodings (pitch, interval, duration) to multi-feature combinations that incorporate octave or duration with pitch. We quantify how these representational choices reshape network topology and evaluate how efficiently each representation structures information by computing both the resulting entropy rate and the divergence from a perceptual-constraint estimate.
Together, these measures make explicit a trade-off: increasing representational detail can preserve finer aspects of musical structure, but may produce a transition structure that is harder for humans to learn and use, while simpler encodings may sacrifice detail yet yield  transition estimates that remain closer to the empirical statistics \cite{tishby2000information,lynn2020human}.

In addition, by further examining how uncertainty and modeled error are distributed locally across each network representation—and how these patterns vary with piece length—we identify conditions under which random-walk centrality aligns with uncertainty.
By quantifying how topology drives divergences between true transition statistics and a perceptually constrained estimate, our approach both reveals a representation-dependent bias that ideal-access expectation models can overlook~\cite{pearce2012auditory,pearce2018statistical} and points toward incorporating realistic memory limits when relating uncertainty to neural~\cite{cheung2019uncertainty} and behavioral data~\cite{gold2019predictability,mas2025predictive}.

\section{Results}
\label{Sec:res}

\begin{figure}[t!]
\centering
\includegraphics[width=0.45\textwidth]{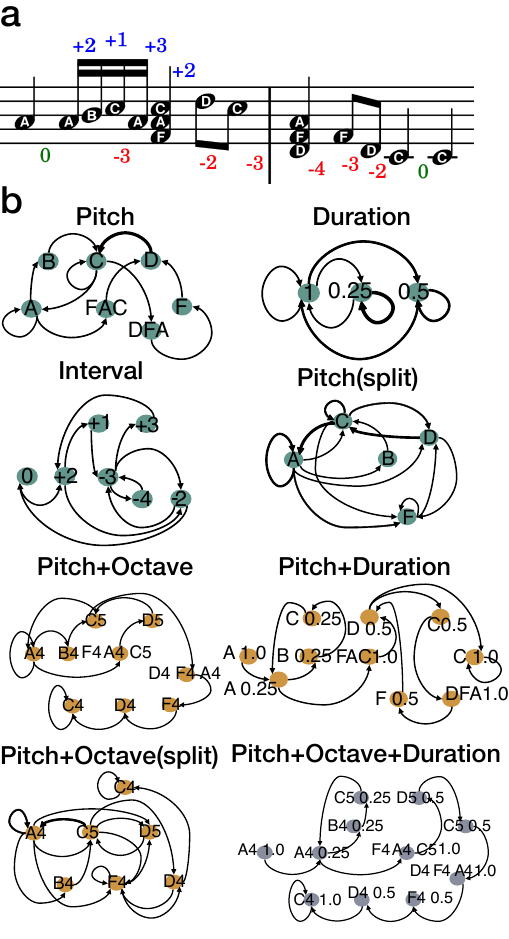}
\caption{\textbf{Networks reconstructed from a musical piece}. (\textbf{a}) A simple composition with highlighted interval changes. (\textbf{b}) Eight distinct network representations. All networks are directed and weighted: edges indicate transitions, and weights correspond to their frequencies.}
\label{fig:0}
\end{figure}

To examine how different levels of musical description shape structural and informational properties, we constructed eight network representations of the same sequences, spanning both established encodings from the literature and controlled feature-level variants introduced for systematic comparison.
We start with minimal, single-feature encodings that expose the basic \textit{building blocks} of musical structure, and then introduce additional features stepwise, allowing us to observe how topological and informational patterns emerge or transform as representational richness increases. In all cases, nodes correspond to unique feature configurations and edges record empirical transitions between consecutive elements in the musical sequence. The eight representations are defined as follows:
\begin{itemize}
   \item \textit{Pitch}: Each node represents the note being played, identified by its pitch class (e.g., C, D, E), without distinguishing the octave. If several notes occur together (a chord), they form a single node \cite{serrà_corral_boguñá_haro_arcos_2012}. 
   \item \textit{Duration}: Each node represents how long a note or chord lasts, expressed as a standardized value relative to a beat.  
   \item \textit{Interval}: Each node represents the distance in semitones between two consecutive notes or chords (see Fig.~\ref{fig:0}a). 
   \item \textit{Pitch+Duration}: Each node represents the note or chord together with its duration. 
   \item \textit{Pitch+Octave}: Each node represents the note or chord together with its octave information. 
   \item \textit{Pitch+Duration+Octave}: Each node represents the note or chord with its duration and octave, combining all three features \cite{ferretti2017,liu_tse_small_2009}. 
   \item \textit{Pitch (split)}: A variation of the \textit{Pitch} model in which chords are split, so that each note is represented as a separate node. Edges are created from the predecessor to each note in the chord, and from each note to the possible successors. 
   \item \textit{Pitch+Octave (split)}: A variation of the \textit{Pitch+Octave} model in which chords are split, with each note and its octave treated as a separate node. As in the previous model, edges connect the predecessor to each note in the chord and from each note to the possible successors~\cite{kulkarni2024information,di2025decoding}.
\end{itemize}
Given a musical piece in MIDI format, we constructed networks by tracking transitions between consecutive musical elements. Each transition forms a directed edge, while the definition of a node depends on the specific model chosen. The resulting networks are both \textit{directed} and \textit{weighted}, with edge weights corresponding to transition counts during the execution of the piece. Figure~\ref{fig:0} illustrates the differences between the different representations. The Supplemental Material~\cite{supplemental_material} further illustrates these differences through network representations of Fryderyk Chopin’s Waltz in E-flat major, Op.~18.

In this work, we focus on piano compositions, which are both musically rich across multiple dimensions and well documented in large, curated corpora. Our corpus combines two sources: the piano-midi.de dataset~\cite{piano-midi}, which contains 268 pieces, and the MSDM dataset~\cite{msdm}, which contains 665 pieces. Together, these datasets cover works ranging from the Baroque period to the 20th century. We preprocessed and merged the two collections, separating the musical scores by left and right hands. Further details are provided in Methods Section~\ref{sec:datasets}.

\subsection{Structural consequences of representational choice}

\begin{figure}[t!]
\centering
\includegraphics[width=0.49\textwidth]{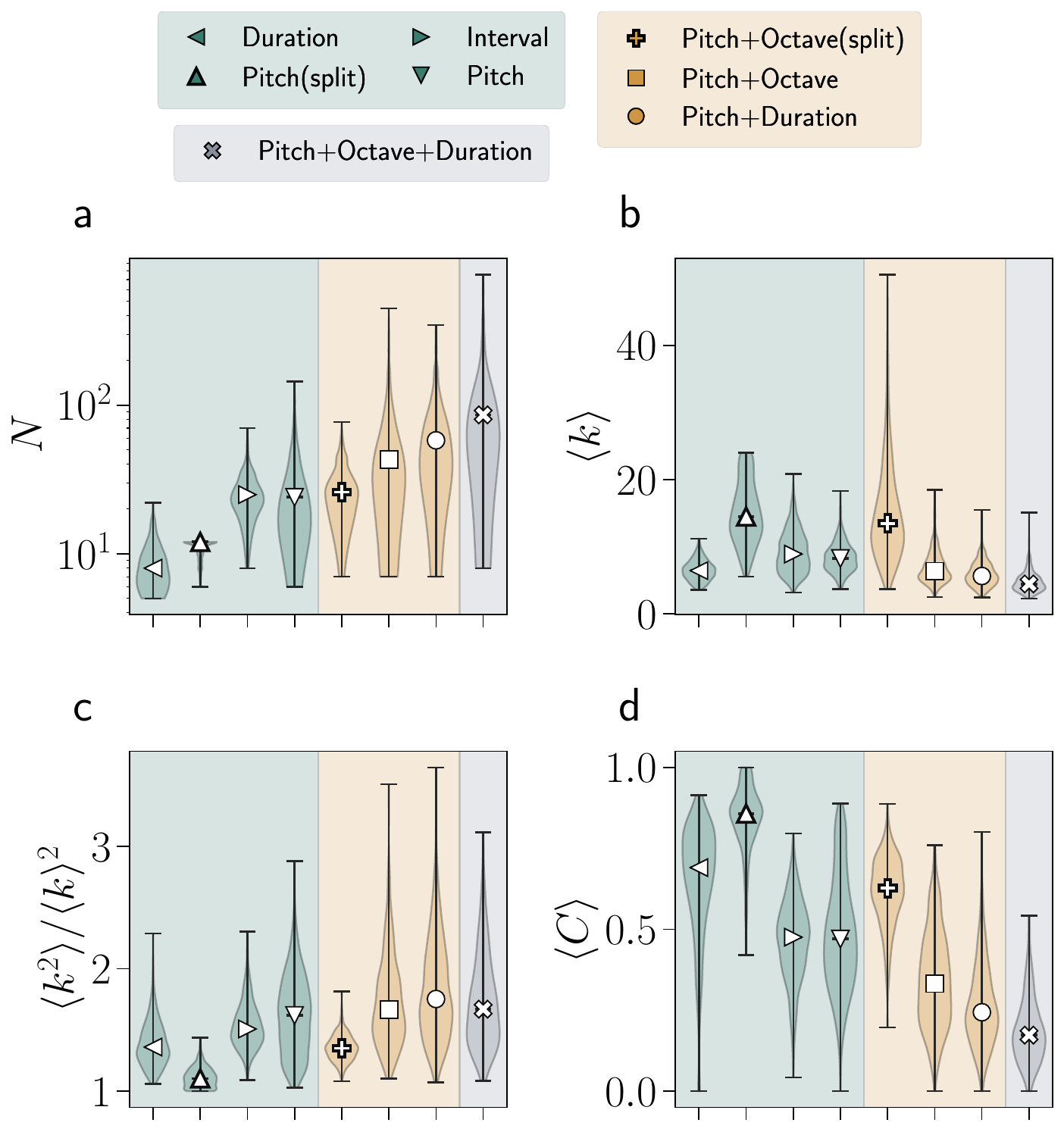}
\caption{\textbf{Topological properties of music network models}. Violin plots showing the distribution of eight network measures: (\textbf{a}) size $N$, (\textbf{b}) average degree $\langle k \rangle$,  (\textbf{c}) degree heterogeneity $\langle k^2 \rangle/\langle k\rangle^2$, (\textbf{d}) average clustering coefficient $\langle C \rangle$, across different models averaged over all musical pieces for the right hand. Models are grouped by the number of musical features used. Each symbol corresponds to an individual model and also highlights the median of each distribution.}
\label{fig:2}
\end{figure}

To assess how representational choices affect network topology, we first focused on four measures that summarize the main structural changes induced by the different event vocabularies: network size, average degree, degree heterogeneity, and clustering coefficient. 
Network size quantifies the number of distinct musical states generated by a representation. The average degree measures the average number of transitions per state, degree heterogeneity captures variability in the degree distribution, and the clustering coefficient measures the extent to which nodes in the graph tend to form tightly connected groups.
These measures provide the structural baseline for the analysis since each representation defines a different set of musical states, and therefore a different transition network. 
Figure~\ref{fig:2} shows these measures for the right-hand networks, while the Supplemental Material~\cite{supplemental_material} reports additional topological measures and the corresponding left-hand results.

The clearest effect of increasing representational detail is an expansion of the state space. 
The most compressed encodings, such as \textit{Duration} and \textit{Pitch(split)}, produce the smallest networks, whereas combined encodings such as \textit{Pitch+Duration} and \textit{Pitch+Octave+Duration} produce the largest networks (Fig.~\ref{fig:2}a). 
For most non-split representations, this expansion is accompanied by a decrease in average degree and clustering coefficient (Fig.~\ref{fig:2}b,d), indicating that transitions become distributed across a larger and sparser set of states rather than concentrated among a small number of recurring event types. 

The \textit{Duration} representation is a limiting case: despite its highly compressed vocabulary, it does not become dense, because the observed duration-to-duration transitions occupy only a restricted subset of the possible state space.
The split representations also depart from the general trend. 
By decomposing simultaneous events into separate note transitions, they create multiple local connections, which keeps them relatively dense and highly clustered even though their state spaces are not among the largest.

At the same time, the expansion is not a simple rescaling of the network. 
Degree heterogeneity increases as additional musical features are incorporated (Fig.~\ref{fig:2}c), showing that richer representations create a stronger contrast between highly connected states and more peripheral states. 
In musical terms, some events become transition hubs that participate in many possible continuations, whereas others occur in more specific sequential contexts. 
Split representations again deviate from this pattern: their local branching increases density and clustering, but it also produces more homogeneous degree distributions than the corresponding non-split representations. 
Thus, the split models should be interpreted as a distinct construction rather than as a simple increase in representational detail.

Together, these results show that representational choice changes more than the number of nodes in the network. Compressed representations concentrate transitions into smaller and denser structures, whereas richer representations expand the state space into sparser and more heterogeneous networks. 
This structural contrast motivates the information-theoretic analyses below, where we move from the topology of each representation to the transition statistics that these topologies support.

\subsection{Informational and perceptual effects of representational choice}

Next, we ask what each representation makes available for prediction. 
Once a musical sequence is encoded as a transition network, each state carries information about what events can follow it. 
However, representations differ in two ways: they differ in how strongly the current state constrains the next event, and, according to recent work \cite{lynn2020abstract,lynn2020human}, they differ in how robust this transition structure is when it is approximated under imperfect access to the sequence.
We therefore analyze the transition networks in two steps. 
First, we quantify the entropy rate of each empirical transition structure. 
This measures, on average, how many continuations remain plausible when the sequence is followed from one state to the next. 
High entropy means that the current encoded state leaves several continuations plausible and is therefore less selective about what can follow; low entropy means that the current state more strongly constrains the next event. 
Thus, entropy is interpreted here as next-state uncertainty: a measure of how predictable the transition structure becomes under a given encoding.
Second, we ask how this same structure changes when it is approximated through a biased transition estimate. 
This captures how empirical transition statistics may be distorted during internalization, before they can support later prediction. 
We model this distortion using a recently proposed framework for how transition probabilities may be internally approximated during abstract sequence learning~\cite{lynn2020abstract}.

First, following prior work on information flow in musical networks~\cite{kulkarni2024information}, we quantify the next-state uncertainty using the Shannon entropy rate of a random walk on the network~\cite{shannon1948mathematical,gomez2008entropy} (see Methods Section~\ref{sec:entropy}),
\begin{align}\label{eq:entropy}
\color{black}
S = \sum_i \pi_i S_i = - \sum_i \pi_i \sum_j P_{ij}\log P_{ij},
\end{align}
where $P_{ij}$ is the probability of moving from state $i$ to state $j$, $S_i$ is the local entropy of the outgoing transitions from state $i$, and $\pi_i$ is the stationary probability of visiting state $i$ in the network. 
The stationary distribution is obtained from the transition matrix $\mathbf{P}$ and reflects how often a diffusive random walker encounters each node in the long run. 
The entropy rate, therefore, captures how next-state uncertainty is dynamically organized across the representation: it is high when frequently visited states have many plausible continuations, and low when frequently visited states have more constrained outgoing transitions. 
In information-theoretic terms, this quantity can be read as the average information produced per step of the encoded sequence; in the present analysis, we use this same quantity to ask how strongly the current encoded state constrains what can follow.

Second, we quantify how this empirical transition structure is altered by the biased transition estimate. 
Using the biased-transition framework introduced by Lynn et al.~\cite{lynn2020abstract,lynn2020human}, we compare the empirical transition matrix $\mathbf{P}$ with a biased estimate $\hat{\mathbf{P}}$. 
\textcolor{black}{The model produces this inferred transition matrix from the empirical transition matrix as}
\begin{align}
\color{black}
\hat{\mathbf{P}} = (1-\eta)\,\mathbf{P}\left(\mathbf{I}-\eta\,\mathbf{P}\right)^{-1},
\end{align}
\textcolor{black}{where $\eta \in [0,1]$ controls the strength of the bias. 
Lower values of $\eta$ keep the estimate closer to the direct one-step transition matrix, whereas higher values increase the contribution of indirect paths. 
This follows from the expansion}
\begin{align}
\color{black}
\left(\mathbf{I}-\eta\,\mathbf{P}\right)^{-1}
=
\sum_{t=0}^{\infty}(\eta \mathbf{P})^t,
\end{align}
\textcolor{black}{which shows that $\hat{\mathbf{P}}$ integrates not only direct transitions, but also higher-order paths of length two, three, or more, each weighted by a decaying factor. 
Consequently, $\hat{\mathbf{P}}$ can be interpreted as a temporally smoothed, biased estimate of the transition structure that would be internalised under imperfect access to the empirical transition statistics.}

We quantify the difference between the empirical and biased transition structures using the Kullback–Leibler (KL) divergence
\begin{align}
\color{black}
D_{KL}(\mathbf{P}\parallel\hat{\mathbf{P}}) =
-\sum_i \pi_i \sum_j P_{ij}\log\frac{\hat{P}_{ij}}{P_{ij}}.
\end{align}
Lower values indicate that the biased estimate remains close to the empirical transition probabilities, whereas higher values indicate that the imposed bias distorts the representation's transition structure more strongly. 
In this sense, the model targets an early step in statistical learning: before transition statistics can support later predictions, they must first be internalised from experience, and this internalisation may itself be biased.

\begin{figure}[t!]
\centering
\includegraphics[width=0.49\textwidth]{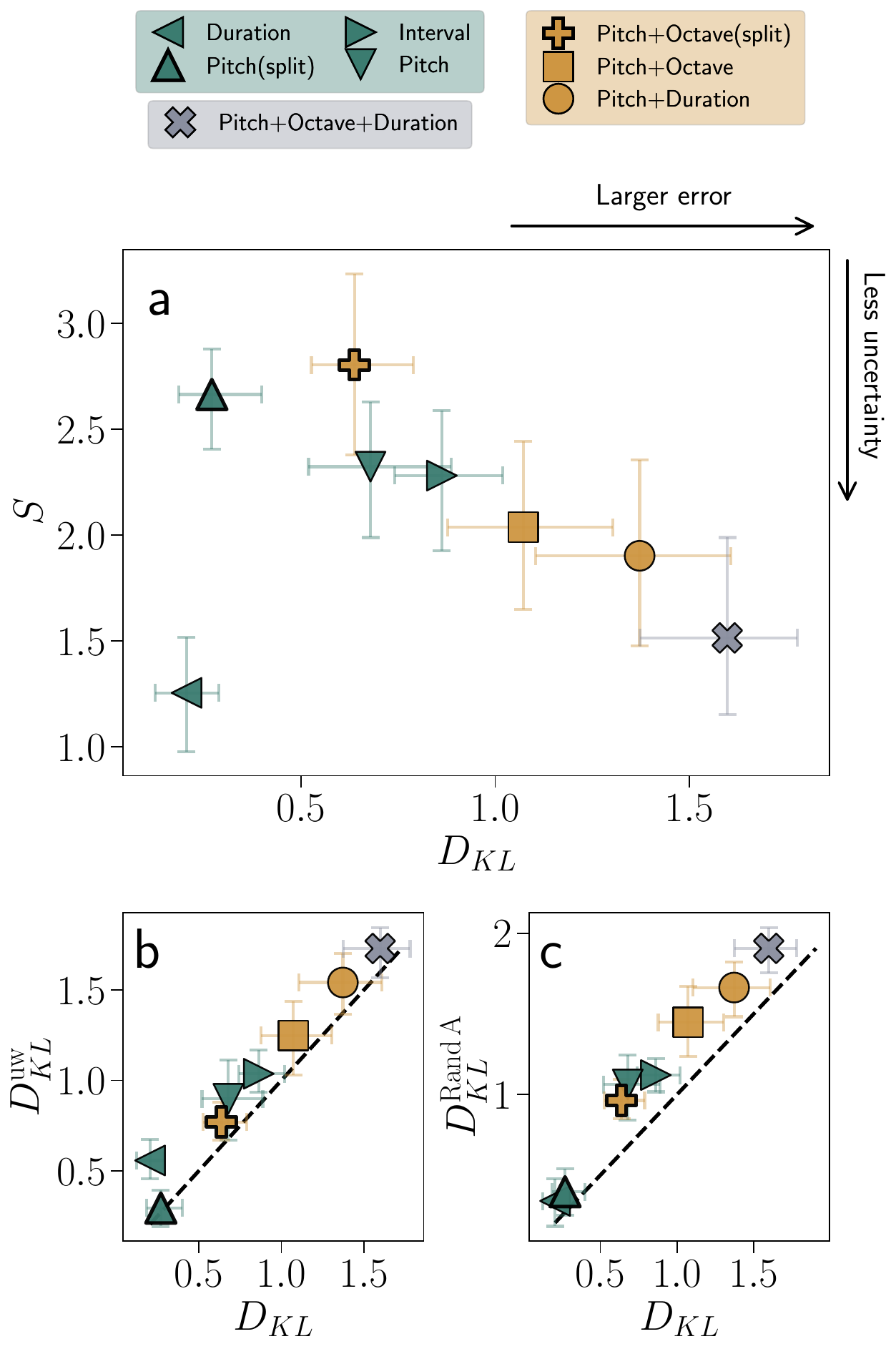} 
\caption{\textbf{Transition uncertainty and robustness to biased internalization across music network representations.} (\textbf{a}) Median $D_{KL}$ versus median $S$. (\textbf{b}) Median Kullback--Leibler divergence of weighted $D_{KL}$ versus unweighted $D^{\mathrm{uw}}_{KL}$ networks. 
(\textbf{c}) Median $D_{KL}$ versus $D_{KL}$ of type-A randomized networks.
Models are grouped by the number of musical features used. Each symbol corresponds to an individual model. Error bars show the interquartile range (IQR) around the median. All panels correspond to right-hand tracks.
}
\label{fig:entropy_kldiv}
\end{figure}

Figure~\ref{fig:entropy_kldiv}a shows that richer musical representations provide a more precise description of what can come next, but this precision makes their transition structure more vulnerable to distortion under the biased estimate. 
Across most representations, increasing the number of encoded musical features lowers next-state uncertainty while increasing internalization error, quantified by $D_{KL}$.
Compressed representations such as \textit{Pitch}, \textit{Duration}, and \textit{Pitch(split)} tend to occupy the low-divergence region: their empirical transition structures remain relatively close to the biased transition estimate. 
However, these representations also encode musical events coarsely, and in several cases retain relatively high entropy because many different continuations are collapsed onto the same state space. 
In contrast, richer representations such as \textit{Pitch+Duration} and \textit{Pitch+Octave+Duration} preserve finer distinctions between events. 
This added specificity tends to lower entropy by making transitions more constrained, but it also increases $D_{KL}$, indicating that the resulting transition structure is more strongly reshaped by the biased estimate.

The split representations should be interpreted separately from this ordering.
By decomposing simultaneous events into separate note transitions, they inflate local branching and introduce additional short-range paths that are not present in the corresponding non-split representations.
This effect is most compressed in \textit{Pitch(split)}, where the state space is bounded by the twelve pitch classes.
As a result, the network can become highly connected within a very small vocabulary, yielding high next-state uncertainty while leaving relatively little detailed transition structure for the biased estimate to disrupt.
\textit{Pitch+Octave(split)} shows the same construction effect in a larger state space: its position in Fig.~\ref{fig:entropy_kldiv}a therefore reflects not only increased representational detail, but
also the specific way in which chords are expanded into sequential transitions.

This pattern reveals a representational trade-off: richer vocabularies make continuations more specific, but their sharper transition profiles are more susceptible to distortion under biased estimation. 
Compressed vocabularies are more robust partly because they collapse distinctions that may be musically meaningful. 
The \textit{Duration} representation is a limiting case, where low entropy and low divergence mainly reflect a very restricted state space.

The bottom panels of Fig.~\ref{fig:entropy_kldiv} test two aspects of this interpretation. 
First, Fig.~\ref{fig:entropy_kldiv}b compares the weighted networks used throughout the main analysis with an unweighted version of the same network skeletons. 
In the weighted networks, transition probabilities reflect how often each transition occurs in the musical sequence. 
In the unweighted version, only the presence or absence of a transition is retained, so all outgoing transitions from a state are treated equally. 
This comparison, therefore, asks whether transition frequencies help preserve the empirical transition structure under the biased estimate, beyond the topology of possible transitions alone.
The unweighted networks show systematically higher $D_{KL}$ than the weighted networks, indicating that empirical transition frequencies reduce the distortion introduced by the biased estimate. 
Repeated transitions, therefore, do not merely add regularity in a trivial sense; they help the biased estimate remain closer to the original empirical transition structure. 
The same pattern is observed for the left hand, as reported in the Supplemental Material~\cite{supplemental_material}.

Second, Fig.~\ref{fig:entropy_kldiv}c asks whether the divergence observed in the real musical networks is specific to their empirical organization. 
We compute $D_{KL}$ for each real network by comparing its empirical transition matrix with its biased estimate, and repeat the same procedure for degree- and out-strength-preserving randomized counterparts (type-A randomization). 
These null models preserve local node-level constraints but disrupt the empirical arrangement of transitions across the network. 
The comparison tests whether divergence depends only on vocabulary size and local connectivity, or also on the global organization of the musical transition structure. 
Real networks show lower $D_{KL}$ than their randomized counterparts, indicating that the empirical placement of transition weights and stationary visitation probability makes the transition structure less vulnerable to distortion by the biased estimate. 
Additional null-model analyses, including randomizations of edge weights on the same network skeleton, are reported in the Supplemental Material~\cite{supplemental_material}.

Together, these results show that representational choice affects both the uncertainty of the empirical transition structure and its robustness to biased internalization. 
Thus, the trade-off is between simplified transition structures that are robust but less specific, and more precise transition structures that are richer as sequential descriptions but more costly to internalize. 
Event representation should therefore not be treated as a neutral preprocessing step: it determines which transition statistics are available for later learning or prediction, and how robust those statistics are during internalization.

\subsection{Stationary flow separates uncertainty and divergence across visited states}

The trade-off between entropy and KL divergence is a global property of each representation. 
However, both quantities are obtained by averaging local node-level values with the stationary distribution (see Eqs.\ref{eq:entropy},\ref{eq:kl_methods}).
The stationary probability of a node indicates how often that state is visited in the long-run dynamics of the network, and therefore how strongly it contributes to the global measures. This means that the same set of local transitions can produce different global values depending on whether uncertainty or divergence is associated with high- or low-visitation states. We therefore next ask how local uncertainty and local divergence are distributed across the stationary visitation profile of each musical network.

\begin{figure}[t!]
\centering
\includegraphics[width=0.49\textwidth]{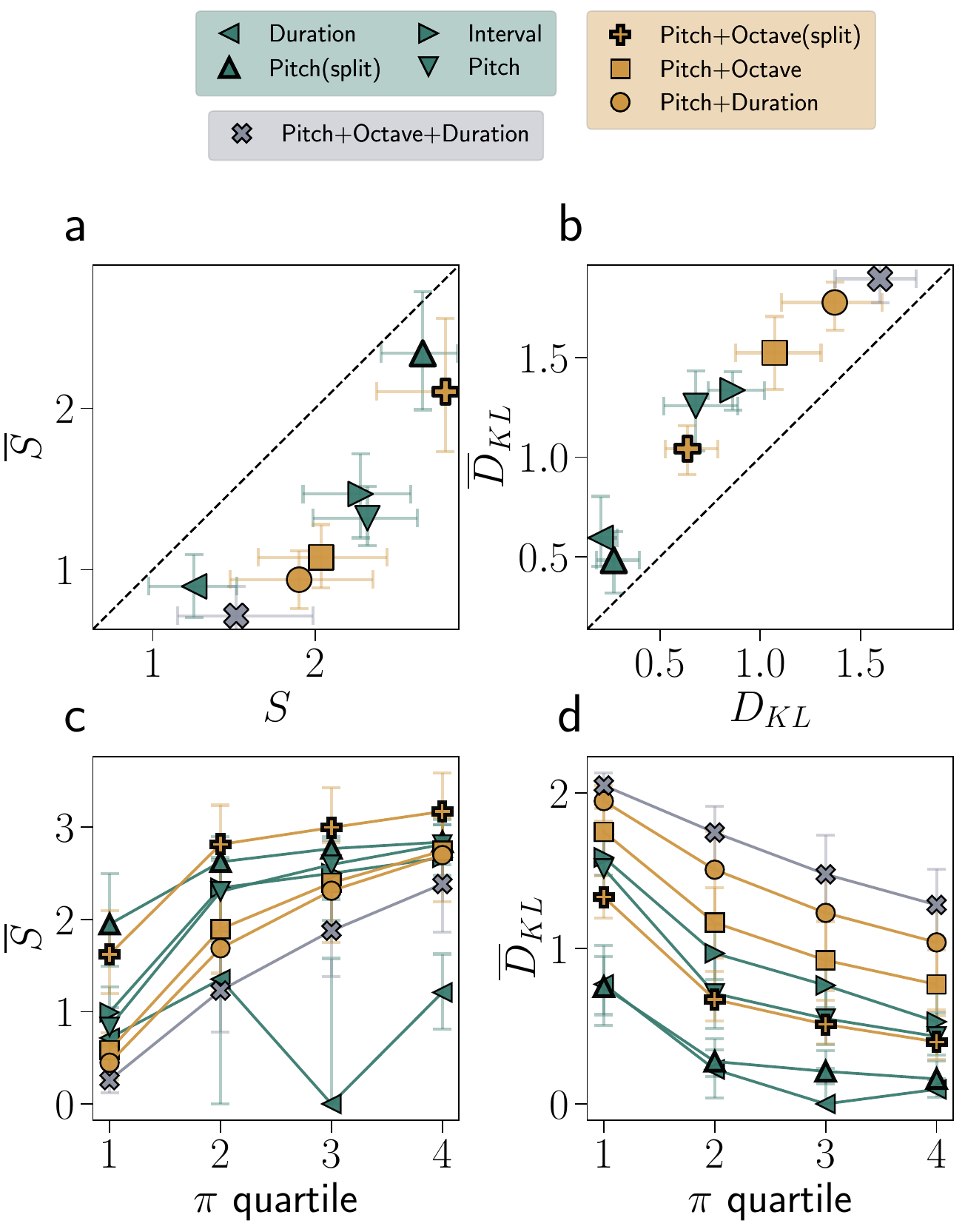} \caption{\textbf{Global and local alignment of uncertainty and inference in music networks.} 
(\textbf{a}) Median entropy $S$ versus mean node-level entropy $\bar{S}$. 
(\textbf{b}) Median KL divergence $D_{KL}$ versus node-averaged KL divergence $\bar{D}_{KL}$. 
(\textbf{c}) Average node entropy ($\bar{S}$) across $\pi$ quartiles.(\textbf{d}) Node-averaged KL divergence ($\bar{D}_{KL}$) across $\pi$ quartiles.
Points show medians, and error bars indicate the interquartile range. All panels correspond to right-hand tracks. Models are grouped by the number of musical features used. Each symbol corresponds to an individual model.}
\label{fig:by_lengths}
\end{figure}

To do so, we compared the entropy rate $S$ computed above with the uniform node average $\bar{S}=\frac{1}{N}\sum_i S_i$. The uniform average treats all states equally, whereas $S$ places greater weight on states visited more often by the random walk. Across representations, $S$ is consistently larger than $\bar{S}$ (Fig.~\ref{fig:by_lengths}a). This shows that high-visitation states tend to have higher local transition uncertainty than the average state. The global entropy rate is therefore determined not only by how uncertain local transitions are, but also by where this uncertainty is located along the stationary distribution.

We then applied the same comparison to the KL divergence. $D_{KL}$ is consistently lower than the uniform node average $\bar{D}_{KL}$ (Fig.~\ref{fig:by_lengths}b).
Thus, states that contribute most strongly to the long-run dynamics tend to be those whose transition probabilities are less distorted by the biased transition estimate. Conversely, states with larger divergence tend to have a lower stationary visitation and therefore contribute less to the global divergence. This also helps explain why empirical transition weights reduce $D_{KL}$ in Fig.~\ref{fig:entropy_kldiv}b: weights do not only define local transition probabilities, but also shape the stationary distribution, increasing the contribution of states whose transition structure is more robust to the bias.

Together, these two comparisons show that uncertainty and divergence are distributed asymmetrically across the stationary distribution of musical networks.
Frequently visited states tend to carry more uncertainty but lower divergence under the modeled perceptual constraints, whereas less frequently visited states tend to carry less dynamical weight and higher divergence. 
From the perspective of biased internalization, this means that the transitions encountered most often are also those whose empirical probabilities are more closely preserved by the biased estimate.

A more local view helps to specify how this organization varies across the stationary distribution. We grouped the nodes into quartiles according to their stationary visitation probability and examined local entropy and local KL divergence within each quartile. The highest-visitation quartiles carry the highest local uncertainty and the lowest local divergence, whereas the lowest-visitation quartiles tend to show the opposite pattern (Fig.~\ref{fig:by_lengths}c,d). Because the stationary distribution is itself highly skewed, global entropy and global divergence are shaped primarily by the subset of states most frequently visited by the random walk rather than by the full state space uniformly. The Supplemental Material~\cite{supplemental_material} reports the corresponding stationary-distribution analyses and left-hand results.

\subsection{Piece length strengthens visitation-based organization of uncertainty and divergence}

Having shown that uncertainty and divergence vary across stationary visitation levels, we next ask whether this visitation-based organization depends on the amount of musical material available in a piece. Longer pieces contain more transitions, and may therefore provide more opportunities for frequently visited and rarely visited states to acquire different informational roles.

To quantify this effect, we defined two visitation-contrast measures by comparing each stationary-weighted quantity with its corresponding uniform node average. For entropy, we used the entropy visitation contrast,
\begin{align}
C_S = S - \bar{S},
\end{align}
which is positive when frequently visited states have higher local entropy than the average state. For KL divergence, we used the divergence visitation contrast,
\begin{align}
C_{KL} = D_{KL} - \bar{D}_{KL},
\end{align}
which is negative when frequently visited states have lower KL divergence than the average state. Thus, both quantities are defined as stationary-weighted minus uniform averages, but their expected signs reflect the asymmetric organization observed above: uncertainty is enriched in frequently visited states, whereas divergence is reduced in frequently visited states.

\begin{figure}[t!]
\centering
\includegraphics[width=0.45\textwidth]{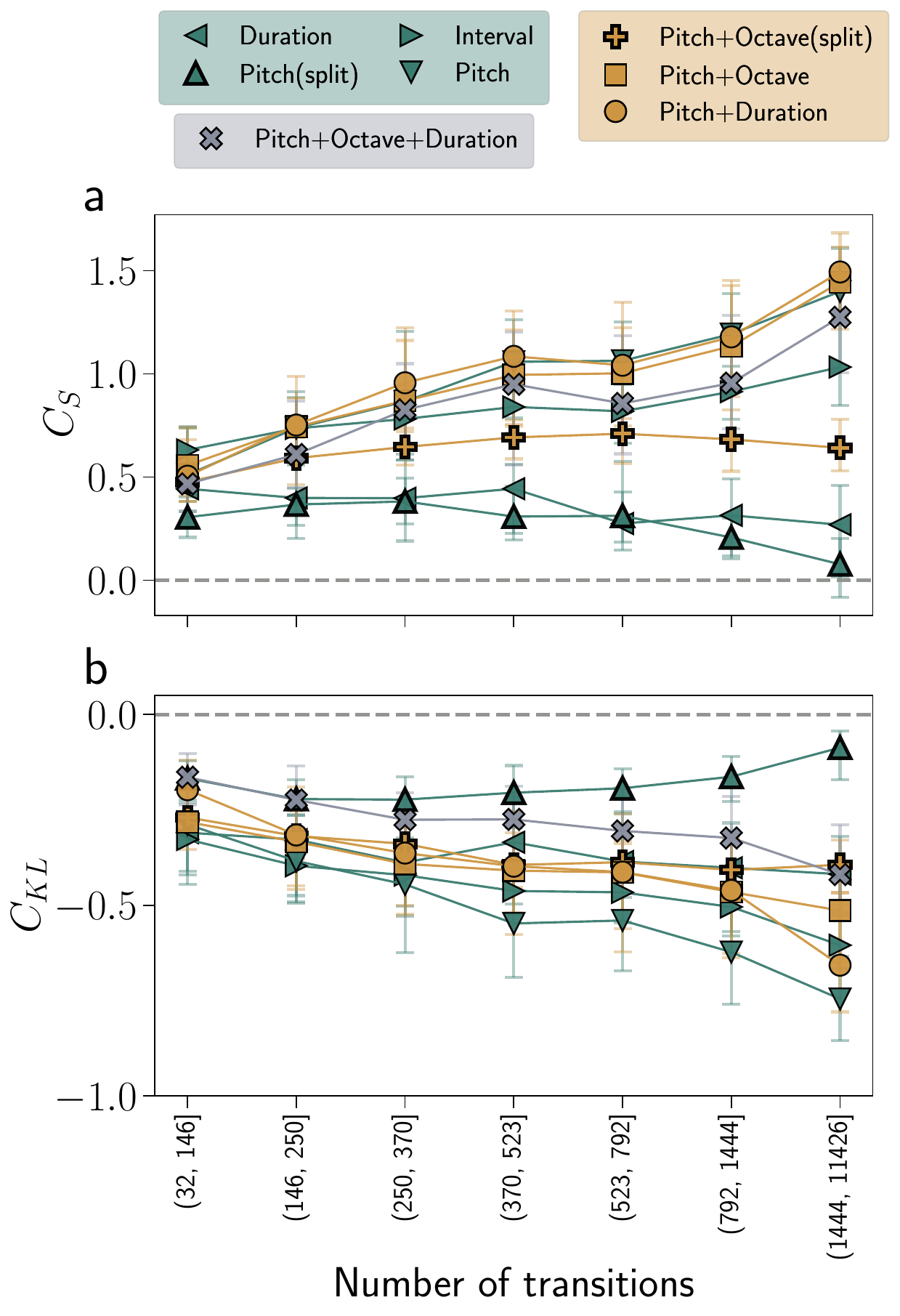} 
\caption{\textbf{Impact of piece length on the organization of uncertainty and inference in music networks.}
(\textbf{a}) Entropy-based visitation contrast $C_S$ as a function of transition count.
(\textbf{b}) Divergence-based visitation contrast $C_{KL}$ as a function of transition count.
Transition counts are divided into seven equal-sized bins. Points indicate medians, and error bars show the interquartile range. All panels correspond to right-hand tracks.}
\label{fig:by_transtion_count}
\end{figure}

As the number of transitions in a piece increases, the two visitation-contrast measures move in opposite directions (Fig.~\ref{fig:by_transtion_count}). The entropy visitation contrast $C_S$ becomes more positive, indicating that longer pieces increasingly concentrate stationary visitation on states with richer local branching (Fig.~\ref{fig:by_transtion_count}a). In contrast, the divergence visitation contrast $C_{KL}$ becomes more negative, indicating that longer pieces increasingly concentrate visitation on states whose empirical transition probabilities remain closer to the biased transition estimate (Fig.~\ref{fig:by_transtion_count}b). Thus, the effect of length is not only to add more transitions or enlarge the state space. Rather, with more musical material, the network develops a stronger visitation-based separation between where uncertainty is carried and where distortion under the bias is concentrated. The corresponding left-hand analysis is reported in the Supplemental Material~\cite{supplemental_material}.

This length dependence differs across representations. In highly compressed or split representations, the trends are weaker and less consistent: the state space is either too small or too locally constrained for a strong visitation-based separation to emerge. In richer multi-feature representations, by contrast, increasing piece length produces a clearer increase in $C_S$ and a clearer decrease in $C_{KL}$. This indicates that the organization observed in Fig.~\ref{fig:by_transtion_count} becomes stronger when the representation contains both sufficient descriptive detail and enough transitions to support differentiated state roles. It also helps explain why these representations differ from the randomized networks discussed above: the empirical sequence does not merely populate a vocabulary, but organizes transition frequencies so that frequently visited and rarely visited states play different informational roles.

Overall, these results explain how the entropy--divergence trade-off develops within individual pieces. Representational detail determines the available state space, while piece length shapes how strongly uncertainty and divergence are separated across the stationary distribution. Thus, longer pieces produce more structured informational landscapes, in which frequently visited states carry greater uncertainty while remaining comparatively less distorted under the biased transition estimate.

\section{Discussion}
\label{sec:discussion}

Representational choice plays a decisive role in the structural properties of musical networks. Our results show that the number and grouping of features—and whether musical events are split or not—substantially influence the resulting topology. Properties often associated with musical corpora, such as small-world organization~\cite{ferretti2017,liu_tse_small_2009}, heavy-tailed degree distributions, or Zipf- and Heaps-like scaling~\cite{manaris2005zipf,zanette2006zipf}, emerge only under certain representational levels and are not inherent to the raw data alone. More detailed, multi-feature representations tend to produce markedly heterogeneous and skewed structures, whereas compressed vocabularies attenuate these patterns. These findings show that topological regularities in music depend on the definition of the underlying symbolic space.

Building on this, our results frame representational differences as a trade-off between descriptive richness and robustness to distortion under perceptual constraints. 
Here, descriptive richness denotes the number of musically relevant distinctions preserved by an encoding. 
This determines how large and how fine-grained its event vocabulary, or state space, is. 
A richer encoding typically makes one-step transitions more specific and more locally predictable, i.e., it lowers next-state uncertainty. 
However, it can also make the transition structure more vulnerable to distortion when transition statistics are estimated under perceptual constraints models \cite{lynn2020abstract}. 
By contrast, robustness to perceptual distortion refers to how closely the constrained transition estimate remains to the empirical transition structure. 
This captures a possible bias in the internalization of transition statistics, rather than the full process by which listeners form predictions.
Compressed representations tend to yield lower distortion, suggesting that their transition topology is less altered by the constrained estimate. 
By contrast, richer representations that preserve finer details reshape the state space and introduce more specific transition structures  that are harder to estimate, increasing the mismatch between empirical and constrained transition estimates. 
This framing also clarifies how our interpretation relates to previous work on information processing in complex and musical networks. 
Lynn et al.~\cite{lynn2020human} and Kulkarni et al.~\cite{kulkarni2024information} use entropy rate and divergence to characterize how efficiently transition structures communicate or process information. 
We use the same formal quantities but emphasize their meaning from the perspectives of prediction and internalization. 
A high entropy rate means that the current encoded state leaves many continuations possible. 
In information-theoretic terms, this corresponds to high information per step, but predictively it also implies that the state is less selective: knowing the current state does not strongly constrain what can come next. 
Such diffuse transition structures can be easier for the biased estimate to approximate, because many alternatives are already present in the empirical transition matrix. 
Conversely, a lower entropy rate can arise when a richer representation defines more specific states with more constrained continuations. 
These states are more useful for prediction because they say more about what is likely to follow, but their sharper transition profiles are also more vulnerable to distortion when indirect paths are mixed into the biased estimate. 
The increase in $D_{KL}$ for richer encodings should therefore not be interpreted as a cost of reduced information, but as the cost of internalizing a more specific and more predictive transition architecture.
This echoes principles from information bottleneck theory~\cite{tishby2000information} and minimum-description-length formulations~\cite{rissanen1983universal}, in which representations balance fidelity against cognitive cost. 
In our network framework, this cost becomes explicit: the representation choice determines both the next-state uncertainty induced by the empirical transition matrix and how strongly that transition structure is distorted when the empirical matrix is approximated by a constrained estimate. 
Viewed this way, representational choices and perceptual constraints jointly determine which transition statistics maybe be available for later learning or prediction.

Although our analysis remains descriptive rather than cognitive, it helps specify a step that is often implicit in statistical-learning accounts of musical expectation: the acquisition of transition statistics from experience. Human expectations depend on learning and exposure~\cite{pearce2018statistical,daikoku2021statistical}, and therefore cannot be reduced to the empirical transitions of a single piece. Our results do not model cognition directly, but they show how the transition statistics available for later prediction may already be altered before prediction occurs. State-of-the-art models of musical expectation~\cite{pearce2012auditory} estimate uncertainty from learned transitions, often assuming that the relevant transition statistics are available without distortion. In this work, we use network representations to make explicit how imperfect access to transition statistics may reshape the statistics available for learning, and to quantify how this effect depends on representational choice by measuring the topology-dependent gap between empirical and constrained transition estimates. A natural extension would be to incorporate this kind of biased transition internalization into statistical-learning models of musical expectation, and test whether predictions based on constrained transition estimates better explain behavioral or neural responses than predictions based on ideal access to empirical statistics. Such an extension would also make it possible to ask whether the visitation-based organization identified here relates to affective responses to music, where uncertainty, surprise, and predictability have been linked to engagement and pleasure~\cite{cheung2019uncertainty,gold2019predictability,huron2008sweet}.

Our representational hierarchy was chosen to build on established symbolic viewpoints in music-network research while enabling a controlled progression from coarse to detailed descriptions. 
Other musically informed viewpoints---such as tonality-relative encodings based on scale degree or harmonic function (e.g., tonic/dominant), or learned feature spaces---may yield different structural signatures and should be explored in future work. 
In particular, future work could ask whether combining multiple musical dimensions within a single predictive framework reduces the trade-off observed here, or whether perceptual distortion persists when pitch, duration, harmony, and longer-range context are learned jointly.
Similarly, our focus on classical piano music provided a rich context for examining how complexity grows with representational detail, but the same framework can be applied to other repertoires. Our intuition is that genres with more repetitive or minimalist structures will show weaker differentiation across representational levels, offering a useful contrast to the dynamics observed here.

More broadly, this approach demonstrates how the structural and informational consequences of representational choice can be analyzed in any sequential domain. Music serves as a clear testbed because its feature combinations and hierarchical abstractions are well-characterized, yet the framework is readily transferable to other sequences and systems. 
By tracing how topologies and informational landscapes evolve with representational detail, we gain a principled way to compare how different feature sets shape the organization of sequential information and the robustness of learned transition statistics under constrained access.

\section{Methods}
\subsection{Datasets and Preprocessing}\label{sec:datasets}
We used two publicly available collections of piano music in MIDI format: the piano-midi.de dataset \cite{piano-midi} (268 pieces) and the MSDM dataset \cite{msdm} (665 pieces). After merging, the corpus comprised 933 files spanning works from the Baroque to the 20th century. Some duplicates remain, since different encodings of the same composition may capture distinct transcription choices.

MIDI files were selected only if they included tempo and time-signature metadata to ensure accurate quantization of durations. We further restricted the corpus to files containing exactly two tracks, corresponding to the left and right hands of a piano score. Each track was converted into a directed and weighted network, resulting in two networks for each composition. To ensure a well-defined entropy with a unique stationary distribution, we restrict our analysis to the largest strongly connected component (LSCC) of each network. Moreover, to ensure that the resulting networks faithfully represented the musical pieces, we retained only LSCCs that contained at least 80\% of the nodes, yielding more than 600 networks per model.

During preprocessing, each musical element was encoded according to the chosen network representation (see Section~\ref{Sec:res}). Symbolic features include pitch (frequency, grouped into pitch classes an octave apart), duration (quantized relative to a quarter note = $1.0$), and chromatic interval (pitch distance in semitones between consecutive events). If multiple pitches occurred simultaneously, the highest pitch was used to compute intervals. Descriptive statistics on pitch ranges, note durations, chord distributions, and interval patterns for both datasets, including systematic differences between left and right hands, are provided in the Supplemental Material~\cite{supplemental_material}.

\subsection{Entropy in complex networks}\label{sec:entropy}
The Shannon entropy for a node $i$ can be written as~\cite{gomez2008entropy,kulkarni2024information}
\begin{align}
    S_i = - \sum\limits_{j} P_{ij} \log P_{ij},
\end{align}
where $P_{ij}$ are the entries of the transition matrix $\mathbf{P}$. In the case of directed and weighted networks, $P_{ij} = \omega_{ij}/s^{\mathrm{out}}_i$, where $\omega_{ij}$ is the edge weight and $s^{\mathrm{out}}_i$ is the out-strength of the node $i$. Here we consider only outgoing links from node $i$. In unweighted networks, the out-strength is replaced by the out-degree $k^{\mathrm{out}}_i$.

The entropy of the entire network has the form of Eq.~\ref{eq:entropy}. The stationary distribution $\pi$ satisfies the condition $\mathbf{P} \pi = \pi$ and can be obtained as the normalized eigenvector of $\mathbf{P}$ corresponding to the eigenvalue 1. For the simplest case, i.e., an undirected and unweighted graph, the stationary distribution becomes $\pi_i = k_i/2E$ where $k_i$ is the degree of node $i$ and $E$ is the total number of edges.
It is worth noting that the formula is only applicable when a network has a single strongly connected component. In the preprocessing step, we ensure that we work with such networks. 

\subsection{Human perception model} 
\label{sec:human_perc}
We adopt the human perception model introduced by Lynn et al.~\cite{lynn2020abstract}, which we briefly summarize here for completeness, as it serves as the basis for our perceptual analysis. In this model, the inferred transition probabilities $\mathbf{\hat{P}}$ are expressed in terms of the true transition probabilities $\mathbf{P}$ as
\begin{align}\label{eq:kl_div}
\mathbf{\hat{P}} = (1 - \eta)\, \mathbf{P}\, \left(\mathbf{I} - \eta\, \mathbf{P}\right)^{-1},
\end{align}
where $\eta \in [0,1]$ controls the inaccuracy of perceptual expectations, and $(\mathbf{I} - \eta \mathbf{P})^{-1}$ denotes a \textit{matrix inverse}. This inverse can be expanded as a geometric series,
\begin{align}
(\mathbf{I} - \eta \mathbf{P})^{-1} = \sum_{t=0}^{\infty} (\eta \mathbf{P})^{t},
\end{align}
which reveals that perception integrates not only direct (one-step) transitions but also higher-order paths---two, three, or more steps apart---each weighted by a decaying factor $\eta^{t}$. 
Consequently, $\mathbf{\hat{P}}$ represents a temporally smoothed higher-order estimate of the transition structure. Empirical evidence suggests an average human accuracy parameter of $\eta \approx 0.8$ which we use for our computation \cite{lynn2020human}.

To quantify the discrepancy in the perception of music networks, i.e., the difference between these two probabilities, we use the Kullback-Leibler (KL) divergence defined as:
\begin{align}
\label{eq:kl_methods}
    D_{KL}(\mathbf{P}\parallel\mathbf{\hat{P}}) = -\sum_{i} \pi_i \sum_{j} P_{ij} \log \frac{\hat{P}_{ij}}{P_{ij}}.
\end{align}
The lower the KL divergence, the smaller the difference between $\mathbf{P}$ and $\mathbf{\hat{P}}$. In addition, we define the uniform average of the KL divergence as
\begin{align}\label{eq:kl_div_uniform}
\bar{D}_{KL}(\mathbf{P} \parallel \mathbf{\hat{P}}) = -\frac{1}{N} \sum_{i=1}^{N} \sum_{j} P_{ij} \log \frac{\hat{P}_{ij}}{P_{ij}}.
\end{align}

\section*{Acknowledgments}
We thank Timoteo Carletti and Nicolás Bono Rosselló for helpful comments and feedback on the manuscript. L.~B.~R and R.~J acknowledge support from the Bridge grant funded by Young Researchers of the Complex Systems Society (yrCSS). R.~J. acknowledges support from the fellowship FI-SDUR funded by Generalitat de Catalunya. L.~B.~R acknowledges support from the ARIAC project (No. 2010235), funded by the Service Public de Wallonie (SPW Recherche). M.~A.~S and M.~B. acknowledge support from grant PID2022-137505NB-C22 funded by MCIN/AEI/10.13039/501100011033 and by ``ERDF/EU''. M.~B. acknowledges the ICREA Academia award, funded by the Generalitat de Catalunya.

\section{Data availability}
The network datasets used in this study are available from the sources referenced in the manuscript.

\section{Code availability}

The code used to construct the network representations and reproduce the main analyses is available at \url{https://github.com/llucbono/network_representations}. The repository includes scripts for reading MIDI files from the analysed datasets, extracting the viewpoint sequences used to define each representation, generating the corresponding directed weighted networks for each MIDI track, and reproducing the analyses reported in the main text.

\vfill
\newpage
\bibliography{refs}

\includepdf[pages={{},{},1,2,3,4,5,6,7,8,9,10,11,12,13,14,15}]{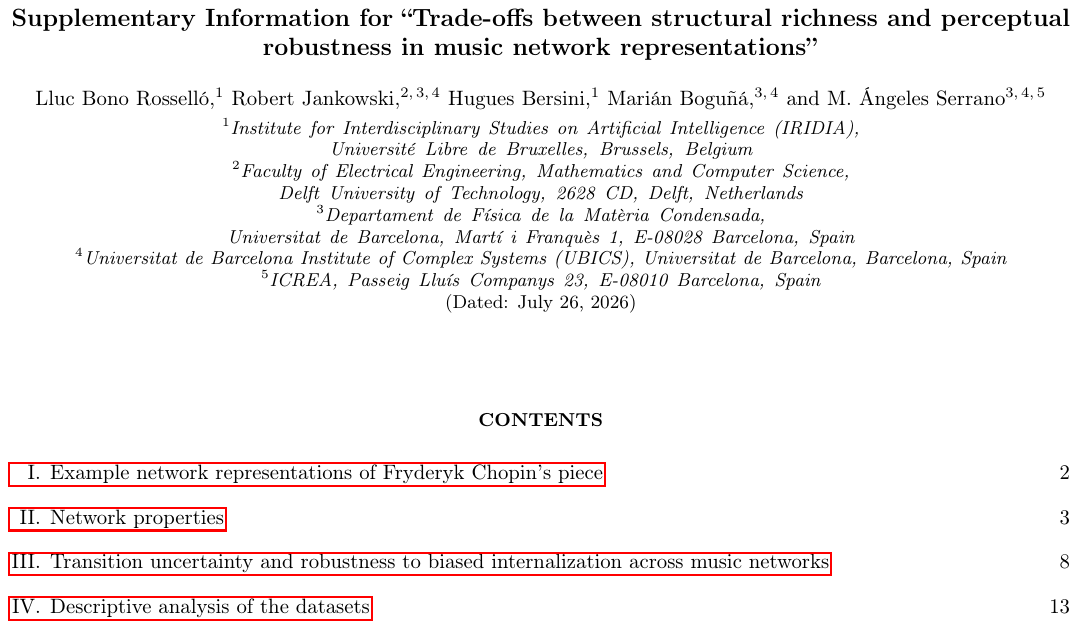}

\end{document}